\documentclass[a4paper]{jpconf}
\usepackage{amsmath,amssymb}
\usepackage{graphicx}
\usepackage{gensymb}
\usepackage{rotating}
\begin{document}
\title{HOC simulation of Moffatt eddies and its flow topology in the triangular cavity flow}

\author{Sougata Biswas$^\dagger$, Jiten C. Kalita$^\ddag$}

\address{Department of Mathematics, Indian Institute of Technology Guwahati, PIN 781039, India}

\ead{$^\dagger$b.sougata@iitg.ernet.in, $^\ddag$jiten@iitg.ernet.in}

\begin{abstract}
In this work, we present HOC simulation of vortices in the triangular cavity for slow viscous incompressible flows by using a recently proposed new paradigm approach for solving Navier-Stokes (N-S) equations. These vortices qualify as Moffatt vortices which are characterized by the computation of common ratios of their sizes and intensities. We further explore topological structures of Moffatt eddies by using crtical point theory which is one of the key concept in the field of topological fluid dynamics.
\end{abstract}
\section{Introduction}\label{intro}
Over the last few years, the formation, evolution, characterization, identification and dynamics of vortices in viscous incompressible internal fluid flows \cite{banks,bru1,bur,ert,gustaf,ghia,haller1,jeo} has been received considerable interest amongst the scientific community. In particular, creeping flows (slow flows) in the neighborhood of the corner of the solid boundary are of customary interest. This flows are referred as ``{\it Stokes flow}" in the existing literature \cite{pant,tasos,white}, which is the linear approximation of the non-linear N-S equations in which non-linear inertia term is not worth considering compared to the linear viscous term.

The existence of an infinite sequence of vortices formed in Stokes flow between two rigid boundaries meeting at a corner is dates back to the pioneering work of Dean and Montagnon \cite{dean} and later theoretically established by Moffatt \cite{moff1, moff2}. A flow visualization experiment by Taneda \cite{taneda} in a V-notch endorsed the existence of these vortices which are now cited as ``{\it Moffatt vortices}" by the fluid dynamicists. The strength and size of each vortex diminishes rapidly as one moves towards the corner. The largest vortex is produced when a cylinder rotates close to the rigid boundaries and a next smaller one is induced by it and the process continues. The orientation of the smaller one is opposite to that largest vortex and the dividing streamlines between them is concave outwards.

Of late there has been a considerable amount of research on the existence of Moffatt vortices in slow viscous incompressible flows on different geometries \cite{kirk,kras1,chetan,malyuga,shankar,shankar1,shtern} established mostly through theoretical studies by eigenvalue analysis.
The theoretical studies on Moffatt vortices are based on the solution of the biharmonic form of the N-S equations for Stokes flow \cite{tasos}
\begin{equation}\label{bhr}
\Delta^2\psi=0
 \end{equation}
in plane polar co-ordinates $(r, \theta)$. Here, $\psi$ is the stream function and $\Delta^2$ is the biharmonic operator. The solution is assumed to be of the form $\displaystyle \psi=r^\nu f_{\nu}(\theta)$, where $\nu$ is any number, real or complex which may be referred as {\it exponent} of the corresponding solution. Then (\ref{bhr}) leads to an equation in $f$ (eigenvalue problem) \cite{shtern,tasos}
 \begin{equation}
 f^{(iv)}+\left\{(\nu+1)^2+(\nu-1)^2\right\}f''
 +(\nu^2-1)^2f=0
 \end{equation}
 which is a linear, homogeneous, fourth-order ordinary differential equation yields a solution of the form \cite{moff1,moff2,shtern,tasos}
 \begin{equation}
 \label{sln}
 f(\theta)=A\sin (\nu-1)\theta+B \cos (\nu-1)\theta+C\sin (\nu+1)\theta+D\cos (\nu+1)\theta,
 \end{equation}
 This $\nu$  was found to be a complex number when the angle between the two planes is sufficiently acute, implying infinite oscillations, i.e. an infinite sequence of counter-rotating eddies as the corner is approached \cite{moff1,moff2}.
 Nonetheless very few numerical studies on this topic are available in the literature \cite{biswas,sbiswas,col,shankar}. Keeping this in mind, we attempt to quantify these vortices in the triangular cavity by the computation of their sizes and intensities.

The objective of the current study is to explore the existence of Moffatt vortices in the triangular cavity by solving the N-S equations numerically at low Reynolds number through a recently developed new paradigm approach by Gupta and Kalita \cite{gupta}. The aforementioned approach has been employed in conjunction with a second order compact scheme which has been mainly used for validating numerical test cases. To the best of our knowledge, this methodology has not been yet utilized for analyzing the existence of Moffatt vortices, in particular in the triangular cavity. We compare our results qualitatively and quantitatively with experimental result \cite{taneda} and theoretical findings \cite{moff1, moff2} available in the literature. We further explore dynamical structures of Moffatt vortices by utilizing rigorous critical point theory from the field of topological fluid dynamics \cite{bakker,delery,hir,wu}. The flow field reveals the presence of ``{\it half-saddles}" on the rigid boundaries leading 
towards the prediction of separation and reattachment; and ``{\it centers}" (dynamical sense) in the interior leading towards the prediction of vortical structures.

The paper is arranged in the following manner: Section 2 deals with the Mathematical description of the problem under consideration, Section 3 with the numerical procedures, Section 4 with results and discussions in which a qualitative study of numerically computing Moffatt eddies has been achieved and flow topology of these eddies has been discussed in details, finally Section 5, end up with the conclusions.

\section{The problem and the governing equations}
\begin{figure}[!ht]
\begin{center}
\includegraphics[height=8cm]{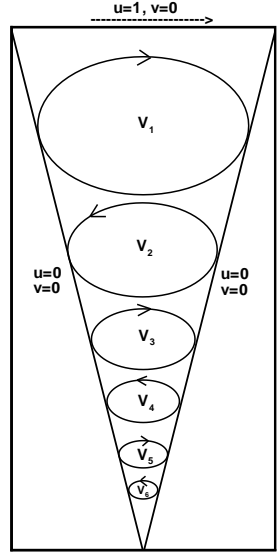}
\caption{Flow configuration in the triangular cavity flow.}\label{config}
\end{center}
\end{figure}

The problem considered here is that of the motion of an incompressible viscous fluid inside a triangular cavity. The cavity is in the shape of an isosceles triangle with an altitude twice the size of its base as in Figure \ref{config} defined inside the rectangular region  $(x, y) \in [0, 1] \times [0, 2]$ in the $xy$-plane.

The equations governing the flow inside the cavity is the famous Navier-Stokes (N-S) equations which in the conventional non-dimensional primitive variable form are given by
\begin{eqnarray}
 \frac{\partial u}{\partial x}+\frac{\partial v}{\partial y}=0, \\
 u\frac{\partial u}{\partial x}+v\frac{\partial u}{\partial y}=-\frac{\partial p}{\partial x}+\frac{1}{Re}\nabla^2u, \\
 u\frac{\partial v}{\partial x}+v\frac{\partial v}{\partial y}=-\frac{\partial p}{\partial y}+\frac{1}{Re}\nabla^2v,\label{eq3}
\end{eqnarray}
where $u$, $v$ are the velocities along the $x$- and $y$- directions respectively, $p$ is the pressure, $Re$ is the non-dimensional Reynolds number,$\nabla^2$ is the Laplacian operator and $U$ is the wall velocity at the lid of the cavity.

Due to the presence of pressure term in the above equations (5-\ref{eq3}), its direct solution is difficult to obtain. To overcome this difficulty, an alternative approach is to employ the streamfunction vorticity formulation $(\psi$-$\omega)$ of N-S equations \cite{white}. 
\begin{eqnarray}
 u\omega_x+v\omega_y =\frac{1}{Re}(\omega_{xx}+\omega_{yy}),\label{s1}\\
 \psi_{xx}+\psi_{yy} = -\omega(x,y)\label{s2},
\end{eqnarray}
with
\begin{equation}
 u=\psi_y,\;v=-\psi_x \quad \mathrm{and}  \quad \omega=v_x-u_y. \label{s3}
\end{equation}

Although this formulation has been quite successful in computing both external and internal incompressible viscous flows, certain difficulty arises as vorticity values along the no-slip boundaries are not specified. Consequently, one needs to look for special measures for obtaining vorticity at the boundaries.

In order to avoid all such difficulties stated above, another alternative formulation can be obtained by the elimination of vorticity term in the $(\psi$-$\omega)$ formulation yields a fourth-order partial differential equation:
\begin{equation}\label{eqq1}
\Delta^2\psi-Re(v\nabla^2u-u\nabla^2v)=0
\end{equation} 
with (\ref{s3}),
where
\begin{equation}\label{eqq3}
\Delta^2\psi= \frac{\partial^4 \psi}{\partial x^4}+2\frac{\partial^4 \psi}{\partial x^2\partial y^2}+\frac{\partial^4 \psi}{\partial y^4}
\end{equation}

By utilizing the equations (\ref{s3}) and (\ref{eqq3}), the equation (\ref{eqq1}) results in the following form:
\begin{equation}\label{eqq4}
\frac{\partial^4 \psi}{\partial x^4}+2\frac{\partial^4 \psi}{\partial x^2\partial y^2}+\frac{\partial^4 \psi}{\partial y^4}-Re\bigg(\frac{\partial^3 \psi}{\partial x^3}+\frac{\partial^3 \psi}{\partial x\partial y^2}\bigg)u-Re\bigg(\frac{\partial^3 \psi}{\partial y^3}+\frac{\partial^3 \psi}{\partial x^2\partial y}\bigg)v=0
\end{equation}
The above form (\ref{eqq4}) is known as the streamfunction-velocity formulation or biharmonic formulation of the N-S equations in 2D set up.

The major advantages of this formulation are: 
\begin{enumerate}
\item[(a)] avoids difficulties to handle the pressure term in the primitive-variables form,
\item[(b)] avoids difficulties for the specification of vorticity boundary conditions in the $\psi$-$\omega$ form,
\item[(c)] iteration procedure involves the only variable $\psi$, and 
\item[(d)] produces high accuracy solutions with little computational cost.
\end{enumerate}
 
The rationale behind choosing the above formulation is to analyse the presence Moffatt eddies numerically in the triangular cavity is that the origin of these vortices lie in the theoretical studies on Stokes flow governed by the biharmonic equation of N-S equations (for details see Section \ref{intro}).

In the triangular cavity under consideration (refer to figure \ref{config} again), all the walls are immobile except the top one in which the non-dimensional $x$-velocity equals to the wall velocity $u=U=1.0$ has been specified. On the immobile walls $x$- and $y$- velocities must be zero due to the presence of no-slip condition. Due to the motion of the top lid, the fluid in contact with it is set into motion and evolves in the interior of the cavity. The flow structure is as depicted in Figure \ref{config} where a sequence of vortices is generated inside the cavity starting from the ones  bigger in size and strength at the top, each one driving the next smaller one and the process continues.
\section{Numerical procedures}
\subsection{The scheme used}
The recently developed second order compact scheme by Gupta and Kalita \cite{gupta} has been used to discretize the equation of the form (\ref{eqq4}). We present here a brief description of their scheme. The aforementioned compact formulation for the 2D steady N-S equations in biharmonic form is given by
\begin{eqnarray}
&&\psi_{i-1,j-1}-8\psi_{i,j-1}+\psi_{i-1,j}-8\psi_{i-1,j}
+28\psi_{i,j}-8\psi_{i+1,j}+\psi_{i-1,j+1}\nonumber\\
&&-8\psi_{i,j+1}+\psi_{i+1,j+1} -3h(u_{i,j-1}-u_{i,j+1}+v_{i+1,j}-v_{i-1,j})-0.5h^2Re\nonumber\\
&&\{v_{i,j}(u_{i+1,j}+u_{i-1,j}+u_{i,j+1}+u_{i,j-1})
-u_{i,j}(v_{i+1,j}+v_{i-1,j}+v_{i,j+1}\nonumber\\
&&+v_{i,j-1})\}=0\label{scm}
\end{eqnarray}
where, $h$ represents step-length on a uniform grid in both $x$- and $y$-directions.

The fourth order central difference formula has been used to approximate the velocities $u$ and $v$ in (\ref{s3}) yields the following formulas:
\begin{eqnarray}
u_{i,j}=\frac{3}{4h}(\psi_{i,j+1}-\psi_{i,j-1})-\frac{1}{4}(u_{i,j+1}+u_{i,j-1}),\\
v_{i,j}=-\frac{3}{4h}(\psi_{i+1,j}-\psi_{i-1,j})-\frac{1}{4}(v_{i+1,j}+v_{i-1,j}).
\end{eqnarray}
\subsection{Associated algebraic systems and its solution}
The algebraic system of equations associated with the compact finite difference scheme (\ref{scm}) can be written as the following matrix equation:
\begin{equation}\label{ase}
M\boldsymbol{\Psi}=\boldsymbol{f}(\psi,u,v)
\end{equation}
where, $M$ is an asymmetric sparse matrix having dimension $mn$ for a grid of size $m \times n$; $\boldsymbol{\Psi}$ and $\boldsymbol{f}$ are $mn$-component vectors. An outer-inner iteration strategy has been employed in order to solve this problem. We solve the equation (\ref{ase}) in the outer iteration cycle by using the biconjugate gradient stabilized method (BiCGStab) \cite{kel,saad} without any preconditioning which constitutes inner iteration cycle. We have used a relaxation parameter $\mu=0.95$ inside both the outer and inner iteration cycles for $\psi$. All of our computations were carried out on on a Pentium core I5 based PC with 3GB RAM. The computations were continued untill  the maximum $\psi$-error between two consecutive outer iteration steps fell bellow $0.5 \times 10^{-9}$.
\subsection{Strategy for boundary conditions}
In the domain $(x, y) \in [0, 1] \times [0, 2]$, we choose the left and right boundary of the triangular cavity in such a way that both the lines corresponding to these boundaries must pass through the available grid points in the computational domain. The boundary conditions for the top wall of the cavity are specified as $u=1,\;v=0$. For other two boundaries the following strategy has been adopted while imposing the boundary conditions as $u=0,\;v=0$ in the computational domain.

For the left boundary: $\quad u_{i,j}=v_{i,j}=0$ in which $i$ varying from $0$ to $(x_{max}-2)/2$ and $j$ varying from $1$ to $(y_{max}-2)$ satisfying $i<\{(x_{max}-1)/2-j/4\}$, where $i,\;j$ stand for $x$-direction index and $y$-direction index respectively. In a similar manner,
for the right boundary: $\quad u_{i,j}=v_{i,j}=0$ in which $i$ varying from $(x_{max}-1)/2$ to $(x_{max}-2)$ and $j$ varying from $1$ to $(y_{max}-2)$ satisfying $i>\{(x_{max}-1)/2+j/4\}$. For the streamfunction-velocity formulation, streamfunction values are set to be zero i.e, $\psi=0$ along all the three boundaries of the cavity.
\section{Result \& Discussions}
In the following, we present a detailed qualitative discussion on Moffatt eddies and its dynamical structures in the flow domain. 
\subsection{Simulation of Moffatt eddies: A qualitative study}
The flow in the triangular cavity has been computed on three different space grids of sizes $129 \times 257$, $ 257 \times 513$ and $513 \times 1025$. However we present the result obtained from the finer grid only. The through investigation of literature reveals the study of Moffatt eddies mainly confined at very low $Re$. Keeping this in mind, we choose $Re$ to be of value $1$ in our numerical computation.

\begin{figure}[!ht]
\begin{center}
\includegraphics[height=10cm]{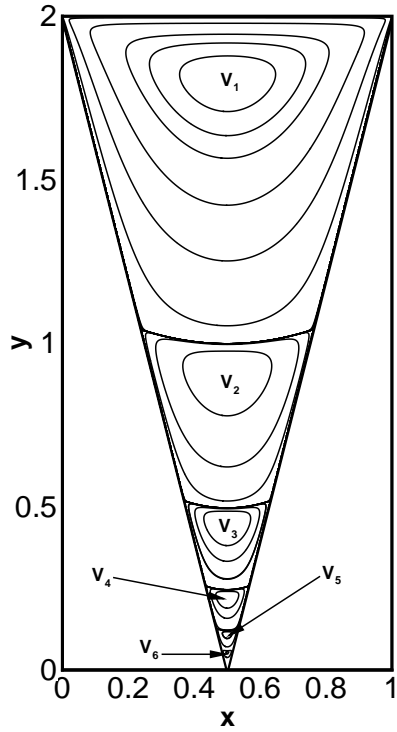}
\caption{Moffatt vortices in the triangular cavity for $Re=1$ on grid of size $513 \times 1025$.}\label{tc_me}
\end{center}
\end{figure}

\begin{figure}[!ht]
 \begin{minipage}{0.5\textwidth}
  \centering
  \includegraphics[height=7.7cm]{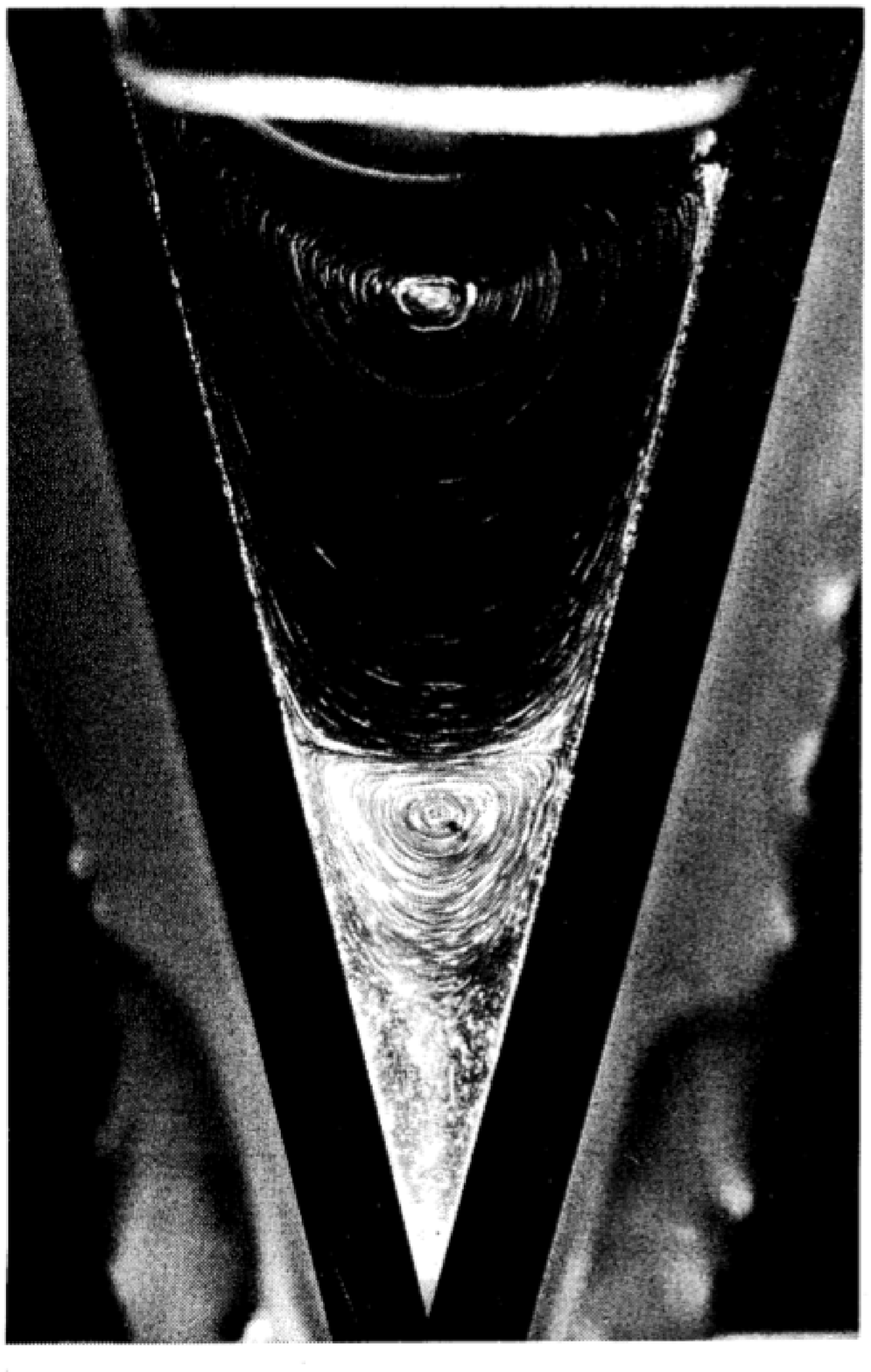}\\
  (a)
 \end{minipage}%
 \begin{minipage}{0.5\textwidth}
  \centering
  \includegraphics[height=5.8cm,angle=-90]{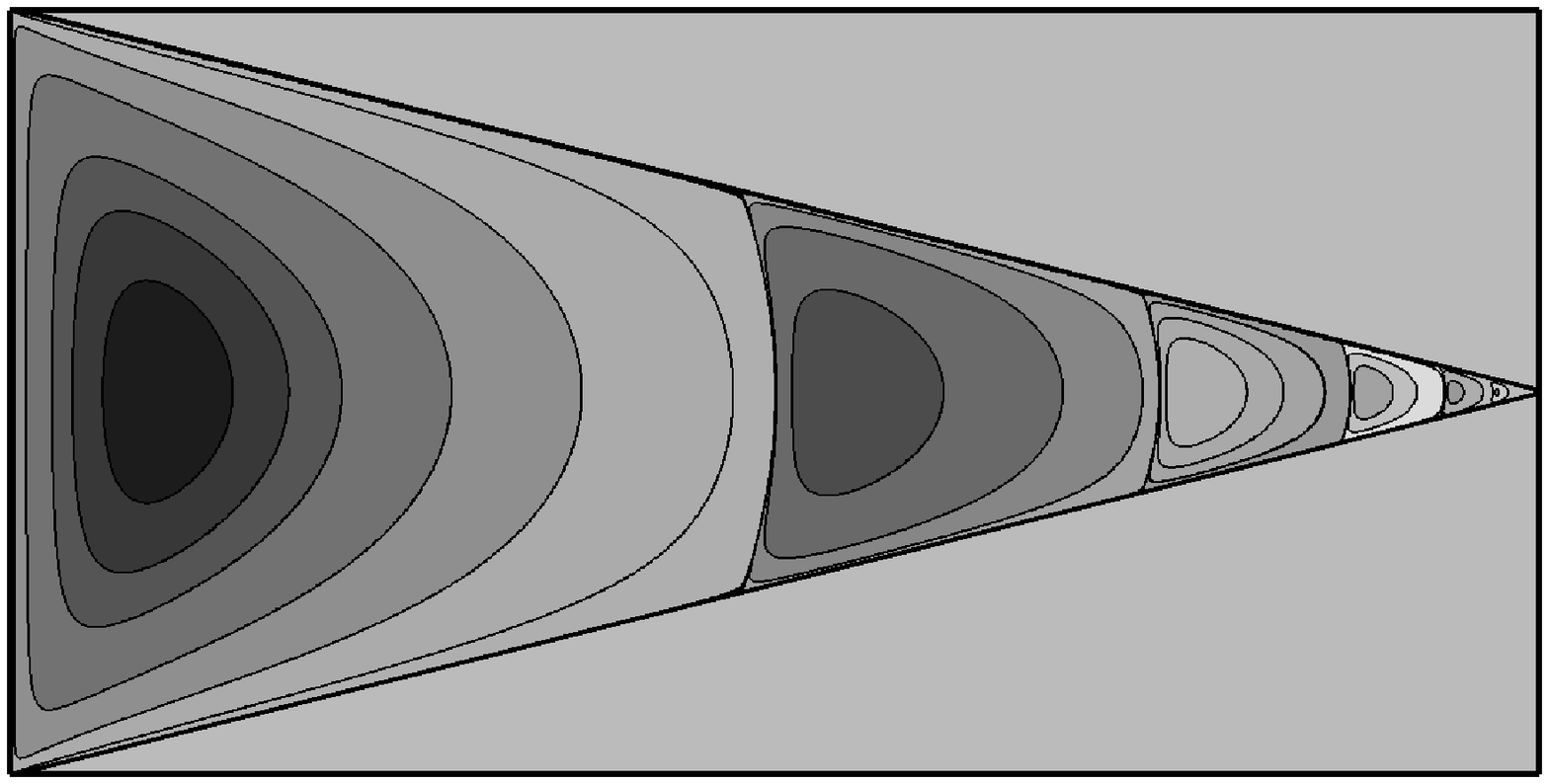}\\
  \vspace{-0.8cm}
  (b)
 \end{minipage}
\caption{ {\sl  (a) Sketch of Taneda's experiment} {\cite{taneda}} {\sl and (b) simulation from our computation}.}\label{ns_exp}
\end{figure}

In Figure \ref{tc_me}, we depict the presence of Moffatt eddies in the triangular cavity. These eddies are stacked from top to bottom in a cascade to the tip. We present them by streamfunction contours and the following nomenclature has been adopted $V_1$, $V_2$, $V_3$ $\cdots$, to mark them to in the sequence sequence as approach to the bottom stationary corner of the cavity. By our computation we are able to trace six members from the sequence of Moffatt eddies. We compare our results with the only experimental result by Taneda \cite{taneda} available in the existing literature. In Figure \ref{ns_exp}, we show comparison between the experimental result and numerically computed result which
shows excellent matching. 

In Table \ref{tab1}, we present the center location, intensity and size of each member in the sequence of Moffatt eddies. The intensity of an eddy is measured by the streamfunction value at its center and size by the vertical distance between its center and tip of the cavity. We observe that the intensity value and size of these eddies decreases rapidly as approaches towards the corner. Moreover the eddies in the sequence fall off a geometric progression having fixed common ratio for both the size and intensity. From our computation, we determine the common ratio of size ($S_R$) and intensity ($I_R$)  as  
$$S_R\bigg(\frac{V_{n+1}}{V_n}\bigg) \approx 0.0012 ~~ \mbox{and} ~~ I_R\bigg(\frac{V_{n+1}}{V_n}\bigg) \approx 0.49,$$ where, $n=1,\;2,\;\cdots ,5$
(For further details see Table \ref{tab2}). All these facts are in accordance with the prediction of the theory of Moffatt \cite{moff1, moff2}.
\begin{table}[!ht]
\begin{center}
\caption{Details of Moffatt eddies in the triangular cavity for $Re=1$ on grid of size $513 \times 1025$}\vspace{0.25cm}
\label{tab1}
\begin{tabular}{cccc}\hline
\textbf{Vortex}
&\textbf{Intensity}
& {\bf Center Location}
&\textbf{Size}
\\ \hline
 $V_1$	& -8.4783 $\times 10^{-1}$  & (0.5000, 1.8027)	& 1.8027\\
 $V_2$	& 1.1159 $\times 10^{-4}$ & (0.5000, 0.9063)	& 0.9063\\
 $V_3$	& -1.3625 $\times 10^{-7}$ & (0.5000, 0.4492)	& 0.4492\\
 $V_4$	& 1.6687 $\times 10^{-10}$ & (0.5000, 0.2227)	& 0.2227\\
 $V_5$	& -2.0694 $\times 10^{-13}$ & (0.5000, 0.1094)	& 0.1094\\
 $V_6$	& 2.6748 $\times 10^{-16}$ & (0.5000, 0.0547)	& 0.0547\\
\hline
\end{tabular}
\end{center}
\end{table}
\begin{table}[!ht]
\begin{center}
\caption{Intensity and geometric ratios of two eddies in succession}\vspace{0.25cm}
\label{tab2}
\begin{tabular}{ccc}\hline
\textbf{Eddy Ratio}
&\textbf{Intensity}
&\textbf{Size}
\\ \hline
 $V_2 : V_1$	& 0.001316  & 0.502709\\
 $V_3 : V_2$	& 0.001221  & 0.495696\\
 $V_4 : V_3$	& 0.001224  & 0.495692\\
 $V_5 : V_4$	& 0.001240  & 0.491468\\
 $V_6 : V_5$	& 0.001292  & 0.499634\\
\hline
\end{tabular}
\end{center}
\end{table}

\begin{figure}[!ht]
\begin{center}
\includegraphics[height=8cm]{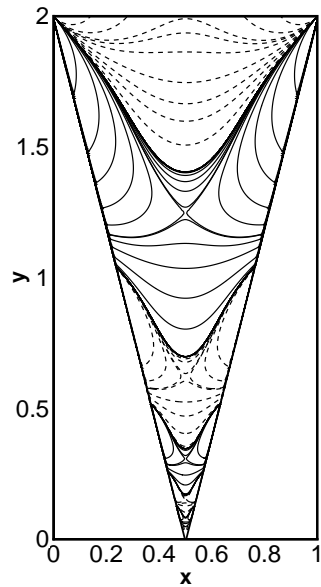}
\caption{Vorticity distribution of Moffatt vortices in the triangular cavity for $Re=1$}\label{med_vort}
\end{center}
\end{figure}

In figure \ref{med_vort}, we present the contour plot of constant vorticity which is defined by $\omega =v_x-u_y$ for $Re=1$. We observe that the vorticity field in the cavity is symmetric about the vertical centerline. It is heartening to note that the characteristics of sequence of eddies and vorticity distribution in the cavity were again in accordance with the analytical predictions of Moffatt \cite{moff1,moff2}.

\subsection{Flow topology of Moffatt eddies}
In this section, we present the dynamical structure of Moffatt eddies by using critical point theory from the field of topological fluid dynamics. Critical point theory furnishes a vocabulary or language to explain complex flow patterns in a simple manner. To understand and interpret the flow patterns whether they have been obtained from experiments by flow visualization technique or from numerical simulations, a broader knowledge of this theory is essential.

A flow field in 2D consist of one or more critical points which are determined by finding the points in which velocity components vanishes simultaneously and the slope of the streamlines become indeterminate. Mathematically, critical points in 2D flows are obtained by finding the fixed points of the trajectories given by
\begin{equation}
\frac{dx}{u(x,y)}=\frac{dy}{v(x,y)}
\end{equation}
where $\vec{V}=u\hat{i}+v\hat{j}$ is the velocity vector in the 2D space $(x,y)$. Different types of critical points such as node, focus, saddle can be observed in the flow field depending on how the streamlines form. The framework behind the characterization of these various critical points can be found in the work of Bakker \cite{bakker} and D\'{e}lery \cite{delery}.

For unique characterization of a flow field it essential to identify all form of critical points correctly. In this study, we have computed the flow inside a triangular cavity  for $Re=1$ by solving the biharmonic form of 2D N-S equations on uniform grids. The computed flow is post processed to identify critical points in the flow field leading to the prediction of separation, reattachment and vortical structures in the flow field.

\begin{figure}[!ht]
\begin{center}
\includegraphics[height=8cm]{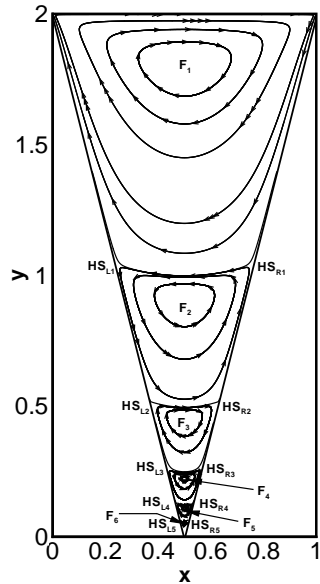}
\caption{Dynamical structure of Moffatt vortices in the triangular cavity for $Re=1$}\label{mes}
\end{center}
\end{figure}

In figure \ref{mes}, we depict streamtraces and velocity vectors in the triangular cavity which reveals the presence of half-saddles on both the left and right boundary of the cavity; the centers (dynamical sense) in the interior of the cavity. In the domain under consideration half-saddles correspond to the prediction of separation points. On the other hand centers correspond to the prediction of vortical structures refereed as Moffatt eddies. We adopt the following nomenclature to present these dynamical structures: for the left boundary the sequence of half-saddles are denoted by $HS_{L1}$, $HS_{L2}$, $HS_{L3} \cdots$; for the right boundary the sequence of half-saddles are denoted by $HS_{R1}$, $HS_{R2}$, $HS_{R3} \cdots$; the centers correspond to Moffatt eddies in the geometric progression are denoted by $F_1$, $F_2$, $F_3 \cdots$.

In Table \ref{tab_crit}, a list of critical points found by our computation has been presented with their spatial location and topological structure. We observe that only two types of half-saddles can be present on the boundary of the flow domain, one leads to separation and the other one to reattachment and thus shed light to the formation of a new eddy. To identify them, we coin the terms ``{\it half-saddle of first kind}" ($HS^1$) for separation points and ``{\it half-saddle of second kind}" ($HS^2$) for reattachment points. In the figure \ref{mes}, on the right boundary of the cavity the half-saddles $HS_{R1}$, $HS_{R3}$, $HS_{R5}$ are of type $HS^1$ while $HS_{R2}$, $HS_{R4}$ are of type $HS^2$. On the other hand, $HS_{L1}$, $HS_{L3}$, $HS_{L5}$ are of type $HS^2$ while $HS_{L2}$, $HS_{L4}$ are of type $HS^1$ on the left boundary. Further on a particular boundary of the flow domain under consideration, the previously mentioned two types of half-saddle point structures appear in an alternate manner. For 
example, on the right boundary $HS_{R1}$ is of type $HS^1$ whereas $HS_{R2}$ is of type $HS^2$ and the process continues. The attributes of these half-saddle point structures determine the orientation of two Moffatt eddies in succession. The dividing streamlines between any two consecutive eddies are connected by these two forms of half-saddle structures. In the figure \ref{mes}, the dividing streamline between $F_1$ and $F_2$ connects the half-saddle structures $HS_{R1}$ and $HS_{L1}$ which are of type $HS^1$ and $HS^2$ respectively. Similar facts can be observed for other members in the sequence as well.

\begin{table}[!ht]
\begin{center}
\caption{Spatial location of critical points and Moffatt eddy structures in the triangular cavity for $Re=1$}\vspace{0.25cm}
\label{tab_crit}
\begin{tabular}{ccc}\hline
\textbf{Critical Points}
&\textbf{Location}
&\textbf{Dynamical Structure}
\\ \hline
 $F_1$	& (0.50120, 1.80294)  & Center\\
 $F_2$	& (0.50006, 0.90533)  & Center\\
 $F_3$	& (0.50000, 0.44948)  & Center\\
 $F_4$	& (0.50000, 0.22265)  & Center\\
 $F_5$	& (0.50000, 0.10977)  & Center\\
 $F_6$	& (0.50000, 0.05362)  & Center\\
 \hline
 $HS_{L1}$ & (0.23832, 1.04077) & Half-Saddle\\
 $HS_{L2}$ & (0.36932, 0.51690) & Half-Saddle\\
 $HS_{L3}$ & (0.43565, 0.25510) & Half-Saddle\\
 $HS_{L4}$ & (0.46712, 0.12586) & Half-Saddle\\
 $HS_{L5}$ & (0.48452, 0.06009) & Half-Saddle\\
 \hline
 $HS_{R1}$ & (0.76149, 1.04119) & Half-Saddle\\
 $HS_{R2}$ & (0.63075, 0.51793) & Half-Saddle\\
 $HS_{R3}$ & (0.56422, 0.25581) & Half-Saddle\\
 $HS_{R4}$ & (0.53281, 0.12623) & Half-Saddle\\
 $HS_{R5}$ & (0.51747, 0.06288) & Half-Saddle\\
\hline
\end{tabular}
\end{center}
\end{table}

\section{Conclusion}
For the first time HOC scheme has been utilized to present a slow viscous incompressible flow simulation in the triangular cavity. The flow features reveal the presence of a sequence of eddies having the properties that their intensities and sizes diminishes rapidly as one moves towards the lower corner of the cavity. We compare our result with experimental result by Taneda \cite{taneda} and analytical results \cite{moff1,moff2} available in the literature and this shows our results are in accordance with them. The vorticity distribution in the cavity also verified the attributes of these eddies. To the best of our knowledge no topological studies of Moffatt eddies in the triangular cavity has not been performed so far. Keeping this mind, we further endeavored to gain some physical insight into the Moffatt eddies in the cavity by using a rigorous topological aspect from the field of topological fluid dynamics. In the flow field, we have identified different forms of critical points such as half-saddle and 
center which uniquely characterizes the flow field.

\section*{References}
\bibliographystyle{plain}

\begin{thebibliography}{60}
\bibitem{banks} D. C. Banks and B. A. Singer. Vortex tubes in turbulent flows: Identification, representation, reconstruction. {\em IEEE Computer Society Press}, In Proceedings of the conference of Visualization, 1994: 132-139.
\bibitem{bakker} P. G. Bakker. Bifurcations in Flow Patterns. {\em PhD thesis, Technical University of Delft, Netherlands }, 1989.
\bibitem{biswas} G. Biswas, M. Breuer and F. Drust. Backward-facing step flows for various expansion ratios at low and moderate Reynolds numbers. {\em Journal of Fluids Engineering, ASME}, 2004; 126(3): 362-374.
\bibitem{sbiswas} S. Biswas and J. C. Kalita, Moffatt vortices in the lid-driven cavity flow, Journal of Physics: Conference Series, 2016; 759: 012081.
\bibitem{bru1} C.-H. Bruneau and M. Saad. The 2D lid-driven cavity problem revisited. {\em Computers and Fluids}, 2006; 35(3): 326-348.
\bibitem{bur} O. R. Burggraf. Analytical and numerical studies of the structure of steady separated flows. {\em Journal of Fluid Mechanics}, 1966; 24(1): 113-151.
\bibitem{col} W. M. Collins and S. C. R. Dennis. Viscous eddies near a 90$\degree$ and a 45$\degree$ corner in flow through a curved tube of triangular cross-section. {\em Journal of Fluid Mechanics}, 1976; 76(3): 417-432.
\bibitem{dean} W.R. Dean and P.E. Montagnon. On the steady motion of viscous liquid in a corner. {\em Mathematical Proceedings of the Cambridge Philosophical Society}, 1949; 45(3): 389-395.
\bibitem{delery} J. D\'{e}lery. Three-dimensional separated flow topology. {\em ISTE Ltd and John Wiley and Sons, Inc.}, 2013.
\bibitem{ert} E. Erturk, T. C. Korke and G. G$\rm {\ddot{o}}$kc$\rm {\ddot{o}}$l. Numerical solutions of 2-D steady incompressible driven cavity flow at high Reynolds numbers. {\em International Journal for Numerical Methods in Fluids}, 2005; 48(7): 747-774.
\bibitem{gustaf} K. E. Gustafson and J. A. Sethian. Vortex Methods and Vortex Motion. {\em SIAM Publications, Philadelphia}, 1991.
\bibitem{ghia} U. Ghia, K. N. Ghia and C. T. Shin. High-Re solutions for incompressible flow using the Navier-Stokes equations and a multigrid method. {\em Journal of Computational Physics}, 1982; 48(3): 387-411.
\bibitem{gupta} M. M. Gupta and J. C. Kalita. A new paradigm approach for solving Navier-Stokes equations: streamfunction-velocity formulation. {\em Journal of Computational Physics}, 2005; 207: 52-68.
\bibitem{haller1} G. Haller. An objective definition of a vortex. {\em Journal of Fluid Mechanics,} 2005; 525: 1-26.
\bibitem{hir} E. H. Hirschel, J. Cousteix and W. Kordulla. Three-dimensional attached viscous flow. {\em Springer-Verlag, Berlin, Heidelberg}, 2014.
\bibitem{jeo} J. Jeong and F. Hussain. On the identification of a vortex. {\em Journal of Fluid Mechanics,} 1995; 285: 69-94.
\bibitem{kel} C. T. Kelley. Iterative methods for linear and nonlinear equations. {\em SIAM Publications, Philadelphia}, 1995.
\bibitem{kirk} E. Kirkinis and S. H. Davis. Moffatt vortices induced by the motion of a contact line. {\em Journal of Fluid Mechanics}, 2014; 746: R3.
\bibitem{kras1} T. S. Krasnopolskaya. Two-dimensional Stokes flow near a corner in a right angle wedge and Moffatt's eddies. {\em Mechanics Research Communications}, 1995; 22(1): 9-14.
\bibitem{chetan} C. P. Malhotra, P. D. Weidman and A. M. J. Davis. Nested toroidal vortices between concentric cones. {\em Journal of Fluid Mechanics}, 2005; 522: 117-139.
\bibitem{malyuga} V. S. Malyuga. Viscous eddies in a circular cone. {\em Journal of Fluid Mechanics}, 2005; 522: 101-116.
\bibitem{moff1} H. K. Moffatt. Viscous and resistive eddies near a sharp corner. {\em Journal of Fluid Mechanics}, 1964; 18(1): 1-18.
\bibitem{moff2} H. K. Moffatt. Viscous eddies near a sharp corner. {\em Archiwum Mechaniki Stosowanej}, 1964; 16(2): 365-372.
\bibitem{pant} R. L. Panton. Incompressible Flow. {\em Wiley, Third Edition}, 2005.
\bibitem{tasos} T. C. Papanastasiou, G. C. Georgiou and A. N. Alexandrou. Viscous Fluid Flow. {\em CRC Press, Boca Raton London New York Washington, D.C.}, 1999.
\bibitem{saad} Y. Saad. Iterative methods for sparse linear systems. {\em PWS Publishing Company}, 1996.
\bibitem{shankar} P. N. Shankar. The eddy structure in Stokes flow in a cavity. {\em Journal of Fluid Mechanics}, 1993; 250: 371-383.
\bibitem{shankar1} P. N. Shankar. Moffatt eddies in the cone. {\em Journal of Fluid Mechanics}, 2005; 539: 113-135.
\bibitem{shtern} V. Shtern, Moffatt eddies at an interface.  Theoretical and Computational Fluid Dynamics, 2014; 28: 651-656.
\bibitem{taneda} S. Taneda. Visualization of separating Stokes flows. {\em Journal of the Physical Society of Japan}, 1979; 46(6): 1935-1942.
\bibitem{white} F. M. White. Viscous Fluid Flow. {\em Tata McGraw Hill, Second Edition}, 1991.
\bibitem{wu} J.-Z. Wu, H.-Y. Ma and M.-D. Zhou. Vortical flows. {\em Springer-Verlag, Berlin, Heidelberg}, 2015.
\end{thebibliography}

\end{document}